\documentclass[a4paper,11pt]{article}

\usepackage{pos}
\usepackage{epsfig,graphicx}
\usepackage{bm,bbm}

\title{Quarkonium production in pNRQCD: the $P$-wave case}

\author{Antonio Vairo}

\affiliation{Physik-Department, Technische Universität München,\\
James-Franck-Str. 1, 85748 Garching, Germany}

\emailAdd{antonio.vairo@tum.de}

\abstract{We review our recent understanding of quarkonium production at $e^+e^-$ and hadronic colliders
  from the point of view of potential non-relativistic QCD, 
  and apply it to $P$-wave charmonium and bottomonium inclusive production.}

\FullConference{
  *** 10th International Workshop on Charm Physics (CHARM2020), ***\\
  *** 31 May - 4 June, 2021 ***\\
  *** Mexico City, Mexico - Online ***
}

\begin{document}
\maketitle

\section{Inclusive quarkonium production in non relativistic EFTs}
\begin{figure}[ht]
\begin{center}
\epsfxsize=8truecm\epsffile{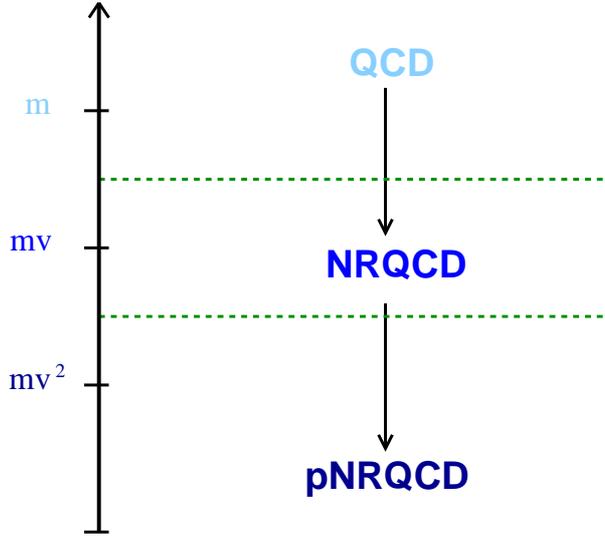}
\caption{Hierarchy of energy scales vs hierarchy of non relativistic EFTs.
\label{fig:scales}}
\end{center}
\end{figure}

The physics of heavy quarkonium (e.g. bottomonium and charmonium)
may be conveniently described in terms of non relativistic effective field theories (EFTs)~\cite{Brambilla:2004jw}.
Indeed, a hierarchy of EFTs can be constructed out of the hierarchy of energy scales characterising any non relativistic bound state:
\begin{equation}
  m \gg mv \gg mv^2,
\end{equation}  
where $m$ is the mass of the heavy quark and $v$ the relative velocity of the heavy quark in the bound state.
The sequence of EFTs that follows from integrating out modes associated with the scales $m$ and $mv$ is shown in figure~\ref{fig:scales}.

Each EFT leads to a factorization of high and low energy contributions in the expression of the observables.
Several quarkonium observables have been computed along the tower of EFTs shown in figure~\ref{fig:scales},
but until recently this has not been the case for quarkonium production cross sections, 
which in the framework of non relativistic effective field theories have been investigated only in non relativistic QCD (NRQCD).
For reviews of such investigations we refer to~\cite{QuarkoniumWorkingGroup:2004kpm,Brambilla:2010cs,Bodwin:2013nua}.
In this contribution, instead, we summarize the work done in the last two years to provide, under some circumstances,  
a description of quarkonium production cross sections also in the ultimate EFT listed in figure~\ref{fig:scales}, i.e.
potential NRQCD (pNRQCD)~\cite{Brambilla:2020xod,Brambilla:2020ojz,Brambilla:2021abf}.

\subsection{NRQCD}
Non relativistic QCD is the EFT that follows from QCD by integrating out quark and gluon modes of energy or momentum of order $m$.
In NRQCD, the inclusive production cross section for a quarkonium $\cal Q$ factorizes into short distance coefficients, $\sigma_{Q \bar Q(N)}$,
encoding contributions from energy scales of order $m$ or larger, 
and into long distance matrix elements (LDMEs), $\langle \Omega | {\cal O}^{\cal Q} (N) | \Omega \rangle$,
encoding contributions from energy scales of order $mv$, $mv^2$ or of the hadronic scale $\Lambda_{\rm QCD}$.
Eventually in NRQCD one writes~\cite{Bodwin:1994jh}:
\begin{equation}
\sigma_{{\cal Q}+X} =  \sum_N \sigma_{Q \bar Q(N)} \langle \Omega | {\cal O}^{\cal Q} (N) | \Omega \rangle\,, 
\label{facNRQCD}
\end{equation}
where in the hadronic case $\sigma_{{\cal Q}+X}$ is the inclusive production cross section of a quarkonium $\cal Q$ with additional light particles~$X$;
in the electromagnetic case we will consider $X=\gamma$.
The state $| \Omega \rangle$ is the QCD vacuum.
The operators ${\cal O}^{\cal Q} (N)$ are four fermion operators that generate and annihilate a heavy quark-antiquark pair.
If the operators project on color singlet states one speaks of color singlet operators,
whereas if they project on color octet states one speaks of a color octet operators.
A more detailed expression of these operators will be given in the following.
The LDMEs for electromagnetic production can be related to those describing quarkonium electromagnetic decay widths.
The NRQCD factorization has been proved for inclusive quarkonium decays, but it has not for hadroproduction.
Nevertheless, a significant effort has been made in the last years
and proofs have been provided at some finite order in perturbation theory~\cite{Nayak:2005rw,Nayak:2005rt,Nayak:2006fm,Bodwin:2019bpf,Zhang:2020atv}.

Equation \eqref{facNRQCD} is an expansion in the strong coupling constant $\alpha_{\text{s}}$,
encoded in the short distance coefficients, and an expansion in $v$, encoded in the LDMEs.
The LDMEs are however poorly known, as they are inherently non perturbative.
They are usually fitted on data, but the inclusion of different set of data in the fit
typically leads to different determinations, none of them able to describe all the data in a satisfactory manner~\cite{Chung:2018lyq,Lansberg:2019adr}.
From the point  of view of figure~\ref{fig:scales}, the LDMEs still contain contributions from the scales $mv$ and $mv^2$. 
Integrating out the scale $mv$ may therefore lead to a further factorization of the LDMEs
and a consequent reduction in the unknown parameters to be fitted to the data.
Hence, moving from NRQCD to pNRQCD could possibly clarify some of the conundrums
plaguing the determinations of the quarkonium production cross sections in NRQCD,
specially in the case of hadroproduction whose color octet matrix elements have no equivalent in the expressions of the quarkonium decay widths.

\subsection{Strongly coupled pNRQCD}
Potential NRQCD is the EFT that follows from NRQCD by integrating out gluon modes of energy or momentum larger than $mv^2$.
We will further assume that 
\begin{equation}
mv^2 \ll \Lambda_{\rm QCD}\,,
\label{strongcoupling}
\end{equation}
which appears to be appropriate for excited (non Coulombic) quarkonium states~\cite{Brambilla:1999xf}.
Potential NRQCD supplemented with the condition \eqref{strongcoupling} is called strongly coupled pNRQCD. 
The factorization formula for the LDMEs in strongly coupled pNRQCD reads at leading order in $v$
(and in the large $N_c$ limit for hadronic color octet matrix elements, $N_c$ being the number of colors)~\cite{Brambilla:2020xod,Brambilla:2020ojz}
\begin{align} 
&\langle \Omega | {\cal O}^{\cal Q}(N) | \Omega \rangle 
=  
\frac{1}{\langle \bm{P}=\bm{0} | \bm{P} = \bm{0} \rangle} \int d^3x_1 d^3x_2 d^3x'_1 d^3x'_2 \, \phi^{(0)}_{{\cal Q}} (\bm{x}_1-\bm{x}_2) 
\nonumber \\ & \hspace{15mm}
\times \left[ - V_{{\cal O}(N)} (\bm{x}_1, \bm{x}_2;\bm{\nabla}_1, \bm{\nabla}_2) 
\delta^{(3)} (\bm{x}_1-\bm{x}'_1) \delta^{(3)} (\bm{x}_2-\bm{x}'_2) \right] \phi^{(0)\,*}_{{\cal Q}} (\bm{x}_1'-\bm{x}_2')\,,
\label{LDMEpNRQCD}
\end{align}
where $| \bm{P} \rangle$ is an eigenstate of the center of mass momentum $\bm{P}$ of the heavy quark-antiquark pair.
We will outline the derivation of \eqref{LDMEpNRQCD} in the following;
the wavefunction $\phi^{(0)}_{{\cal Q}}(\bm{x}_1-\bm{x}_2)$ and the contact term $V_{{\cal O}(N)}(\bm{x}_1, \bm{x}_2;\bm{\nabla}_1, \bm{\nabla}_2)$
are determined by matching the NRQCD LDMEs to pNRQCD.

\subsubsection{Matching the spectrum}
\begin{figure}[ht]
\begin{center}
\epsfxsize=8.2truecm \epsfbox{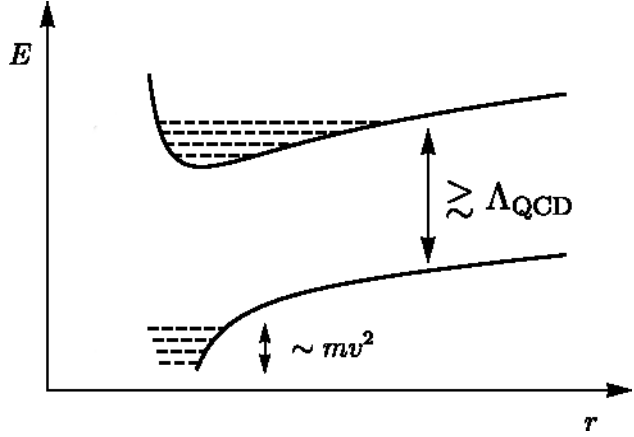}
\caption{Schematic distribution of energy levels induced by the static quark-antiquark potential, which is of the Cornell type, and its first gluonic excitation.
  \label{fig:figgap}}
\end{center}
\end{figure}

Before matching the LDMEs, we need to match the NRQCD Hamiltonian $H_{\rm NRQCD}$~\cite{Brambilla:2000gk,Pineda:2000sz,Brambilla:2003mu}.
The spectral decomposition of $H_{\rm NRQCD}$ in the heavy quark-antiquark sector of the Hilbert space reads
\begin{equation}
  H_{\rm NRQCD} =
  \sum_n \int d^3x_1\,d^3x_2\,|\underline{\rm n}; \bm{x}_1, \bm{x}_2 \rangle \,E_n(\bm{x}_1, \bm{x}_2; \bm{\nabla}_1, \bm{\nabla}_2)\,
  \langle\underline{\rm n}; \bm{x}_1, \bm{x}_2 |\,,
\end{equation}
where $|\underline{\rm n}; \bm{x}_1, \bm{x}_2 \rangle = \psi^\dag(\bm{x}_1) \chi(\bm{x}_2) | n; \bm{x}_1, \bm{x}_2 \rangle$
are orthonormal states made of a heavy quark, $\psi$, a heavy antiquark, $\chi$, and some light degrees of freedom labeled by $n$.
The states $| n; \bm{x}_1, \bm{x}_2 \rangle$ do not contain heavy particles and are also normalized.
The operators $E_n(\bm{x}_1, \bm{x}_2; \bm{\nabla}_1, \bm{\nabla}_2)$ depend on the coordinates, momenta and spin of the heavy quark-antiquark pair.
In the static limit, $E_n(\bm{x}_1, \bm{x}_2; \bm{\nabla}_1, \bm{\nabla}_2)=E_n^{(0)}(\bm{x}_1- \bm{x}_2)$ are the different energy excitations of a static quark-antiquark pair,
with $E_0^{(0)}(\bm{x}_1- \bm{x}_2)$ being the quarkonium static energy.
They may be computed in lattice QCD as a function of the relative distance $\bm{r} = \bm{x}_1-\bm{x}_2$ of the two static sources.
Computations done in the hybrid sector~\cite{Bali:2000vr,Juge:2002br,Capitani:2018rox} suggest that the different static energies $E_n^{(0)}(\bm{r})$,
if not degenerate, are separated at large distances by energy gaps of order $\Lambda_{\rm QCD}$, which is consistent with expectations from non perturbative QCD. 
In the strong coupling regime, $mv^2 \ll \Lambda_{\rm QCD}$,
the energy levels associated to each static energy are expected to distribute schematically as in figure~\ref{fig:figgap}. 
The eigenstates of the NRQCD Hamiltonian in the heavy quark-antiquark sector can be written as 
\begin{equation}
| {\cal Q} (n,\bm{P}) \rangle = \int d^3x_1 d^3x_2 \, \phi_{{\cal Q}(n,\bm{P})}(\bm{x}_1,\bm{x}_2) \, |\underline{\rm n}; \bm{x}_1, \bm{x}_2 \rangle\,,
\label{QnP}
\end{equation}
where the functions $\phi_{{\cal Q}(n,\bm{P})}(\bm{x}_1,\bm{x}_2)$ are eigenfunctions of $E_n(\bm{x}_1, \bm{x}_2; \bm{\nabla}_1, \bm{\nabla}_2)$.
The state $| {\cal Q} (n,\bm{P}) \rangle$ is made of a heavy quark-antiquark pair moving with center of mass momentum $\bm{P}$.

In strongly coupled pNRQCD, the pNRQCD Hamiltonian that describes the low energy dynamics of a color singlet field $S_n$
made of a heavy quark-antiquark pair and light degrees of freedom in a state $n$ reads
\begin{equation}
H_{\rm pNRQCD} = \int d^3x_1 \, d^3x_2 \;  S_n^\dag \,h_n (\bm{x}_1, \bm{x}_2; \bm{\nabla}_1, \bm{\nabla}_2) \,S_n\,.
\end{equation}
The pNRQCD Hamiltonian $h_n$ is obtained by matching the NRQCD energy $E_n(\bm{x}_1, \bm{x}_2; \bm{\nabla}_1, \bm{\nabla}_2)$.
The matching may be performed order by order in $1/m$ by expanding the NRQCD Hamiltonian
and computing the states $|\underline{\rm n}; \bm{x}_1, \bm{x}_2 \rangle$ in quantum mechanical perturbation theory.
At leading order in $v$ we obtain 
\begin{equation}
  h_n(\bm{x}_1, \bm{x}_2; \bm{\nabla}_1, \bm{\nabla}_2) = - \frac{\bm{\nabla}_1^2}{2 m} - \frac{\bm{\nabla}_2^2}{2 m} + V^{(0;n)} (\bm{x}_1 - \bm{x}_2)\,.
\label{eq:hn}
\end{equation}  
The matching fixes the static potential $V^{(0;n)} (\bm{x}_1 - \bm{x}_2)$  to be the static energy $E_n^{(0)} (\bm{x}_1 - \bm{x}_2)$.
In particular,  $V^{(0;0)} (\bm{x}_1 - \bm{x}_2)$ is the quarkonium static energy. 
As a consequence of the matching, the functions $\phi_{{\cal Q}(n,\bm{P})}(\bm{x}_1,\bm{x}_2)$ are eigenfunctions of $h_n$.
The functions $\phi_{{\cal Q}(n,\bm{P})}^{(0)} (\bm{x}_1,\bm{x}_2)$ are the functions $\phi_{{\cal Q}(n,\bm{P})}(\bm{x}_1,\bm{x}_2)$ at leading order in $v$,
i.e. the eigenfunctions of the right-hand side of equation~\eqref{eq:hn}.

\subsubsection{Matching the LDMEs}
We consider, first, LDMEs for electromagnetic production of heavy quarkonium.
LDMEs for electromagnetic production of quarkonia have the form 
\begin{equation}
{\cal O}^{\cal Q} (N) = \chi^\dag {\cal K}_N \psi |{\cal Q}(0,{\bf P})\rangle \langle {\cal Q}(0,{\bf P})| \psi^\dag {\cal K}'_N \chi\,,
\end{equation}
where ${\cal K}_N$ and ${\cal K}'_N$ are generic gauge covariant operators not containing heavy particle fields.

The state $|{\cal Q}(0,{\bf P})\rangle$ describes a quarkonium state with center of mass momentum $\bm{P}$.
This state can be written as in equation~\eqref{QnP} setting $n=0$.
The functions $\phi_{{\cal Q}(0,{\bf P})}(\bm{x}_1,\bm{x}_2)$ are the quarkonium wavefunctions.
At leading order in $v$, they may be approximated by $\phi_{{\cal Q}(0,\bm{P})}^{(0)} (\bm{x}_1,\bm{x}_2)$,
which are the solutions of the Schr\"odinger equation with the static potential $V^{(0;0)} (\bm{x}_1 - \bm{x}_2)$.

The matching conditions for the electromagnetic production contact terms $V_{{\cal O}(N)}(\bm{x}_1, \bm{x}_2;\bm{\nabla}_1, \bm{\nabla}_2)$,
which appear in equation~\eqref{LDMEpNRQCD}, are 
\begin{align} 
& \int d^3x \, \langle \Omega | \left(\chi^\dag {\cal K}_N \psi \right) (\bm{x}) | \underline{\rm 0}; \bm{x}_1, \bm{x}_2 \rangle 
\langle \underline{\rm 0}; \bm{x}'_1, \bm{x}'_2 | \left(\psi^\dag {\cal K}'_N \chi\right) (\bm{x}) | \Omega \rangle 
\nonumber \\ 
& \hspace{5cm}
                = - V_{{\cal O}(N)} (\bm{x}_1, \bm{x}_2;\bm{\nabla}_1, \bm{\nabla}_2)\, \delta^{(3)} (\bm{x}_1-\bm{x}'_1) \delta^{(3)} (\bm{x}_2-\bm{x}'_2)  \,.
\end{align} 
Since we may express the states $|\underline{\rm n}; \bm{x}_1, \bm{x}_2 \rangle$ as series in $1/m$,
the contact terms $V_{{\cal O}(N)}(\bm{x}_1, \bm{x}_2;\bm{\nabla}_1, \bm{\nabla}_2)$ are computed order by order in $1/m$.

Color singlet and color octet operators for hadroproduction of quarkonia have the form respectively
\begin{align}
{\cal O}^{\cal Q} (N_{\textrm{color singlet}}) &= \chi^\dag {\cal K}_N \psi {\cal P}_{{\cal Q}(\bm{P}=\bm{0})} \psi^\dag {\cal K}'_N \chi\,,
  \\
{\cal O}^{\cal Q} (N_{\textrm{color octet}}) &= \chi^\dag {\cal K}_N T^a \psi \Phi_\ell^{\dag ab} (0) {\cal P}_{{\cal Q}(\bm{P}=\bm{0})} \Phi_\ell^{bc} (0) \psi^\dag {\cal K}'_N T^c \chi\,,
\end{align}
where $\Phi_\ell(x)$ is a Wilson line along the direction $\ell$ in the adjoint representation extending from $x$ to $x+l\infty$
required to ensure the gauge invariance of the color octet LDME~\cite{Nayak:2005rw}.
The operator ${\cal P}_{{\cal Q}(\bm{P})}$ projects  onto a state containing a heavy quarkonium $\cal Q$ with momentum $\bm{P}$.
The projector commutes with the NRQCD Hamiltonian (the number of quarkonia is conserved) and therefore is diagonalized by the same eigenstates of the NRQCD Hamiltonian.
It has the form 
\begin{equation}
{\cal P}_{{\cal Q}(\bm{P})} = \sum_{n \in {\mathbb S}} | {\cal Q} (n,\bm{P}) \rangle \langle {\cal Q} (n,\bm{P}) |\,.
\label{eq:P}
\end{equation}
The sum extends over the set ${\mathbb S}$ that contains all states where the heavy quark-antiquark pair forms a color singlet at the origin in the static limit.
This is a necessary condition to produce a quarkonium.

Through equation~\eqref{QnP} the projector ${\cal P}_{{\cal Q}(\bm{P})}$ depends on the wavefunctions $\phi_{{\cal Q}(n,\bm{P})}(\bm{x}_1,\bm{x}_2)$ with $n \in {\mathbb S}$.
At leading order in $v$, they can be approximated by the  wavefunctions $\phi_{{\cal Q}(n,\bm{P})}^{(0)}(\bm{x}_1,\bm{x}_2)$, 
which are solutions of the Schr\"odinger equation with static potential $V^{(0;n)}(\bm{x}_1-\bm{x}_2)$.
The static potential $V^{(0;n)}$ can be computed from the energy exponent associated to a static Wilson loop in the presence of some disconnected gluon fields
selecting the $n$th excitation of the color singlet quark-antiquark pair.
Lattice QCD determinations of $V^{(0;n)}$ for $n \in {\mathbb S}$ and $n \neq 0$ are not available yet.
Nevertheless, the expectation is that disconnected gluon fields produce mainly a constant shift to the potentials, for instance in the form of a glueball mass.
To the same conclusion one arrives by looking at the large $N_c$ limit.
At large $N_c$, the vacuum expectation value of a Wilson loop in the presence of additional disconnected gluon fields factorizes
into the vacuum expectation value of the Wilson loop times the vacuum expectation value of the additional gluon fields
up to corrections of order $1/N_c^2$~\cite{Makeenko:1979pb,Witten:1979pi}.
Since the slopes of the static potentials are all the same for $n \in {\mathbb S}$, we may approximate  
\begin{equation}
  \phi_{{\cal Q}(n,\bm{P})}(\bm{x}_1,\bm{x}_2)  \approx e^{i \bm{P} \cdot (\bm{x}_1+\bm{x}_2)/2} \phi_{{\cal Q}(0,\bm{0})}^{(0)} (\bm{x}_1,\bm{x}_2)\,.
\label{phin-approx}
\end{equation}
This approximation is valid up to corrections of relative order $v^2$ and $1/N_c^2$.
In equation~\eqref{phin-approx} we have also made explicit the leading order dependence on the center of mass momentum,
which is that one of a plane wave; $\phi_{{\cal Q}(0,\bm{0})}^{(0)} (\bm{x}_1,\bm{x}_2)$ depends only on the relative distance $\bm{r}$. 

The matching conditions for the hadroproduction contact terms $V_{{\cal O}(N)}(\bm{x}_1, \bm{x}_2;\bm{\nabla}_1, \bm{\nabla}_2)$,
which appear in equation~\eqref{LDMEpNRQCD}, are for singlet and octet respectively
\begin{align} 
& \sum_{n \in {\mathbb S}} \int d^3x \, \langle \Omega | \left( \chi^\dag {\cal K}_N \psi \right) (\bm{x}) | \underline{\rm n}; \bm{x}_1, \bm{x}_2 \rangle 
\langle \underline{\rm n}; \bm{x}'_1, \bm{x}'_2 | \left( \psi^\dag {\cal K}'_N \chi \right) (\bm{x}) | \Omega \rangle 
\nonumber \\ 
& \hspace{5cm}
    = - V_{{\cal O}(N)} (\bm{x}_1, \bm{x}_2;\bm{\nabla}_1, \bm{\nabla}_2) \, \delta^{(3)} (\bm{x}_1-\bm{x}'_1) \delta^{(3)} (\bm{x}_2-\bm{x}'_2)\,,
\\
  & \sum_{n \in {\mathbb S}}
    \int d^3x \langle \Omega | \left( \chi^\dag {\cal K}_N T^a \psi \right) (\bm{x}) \Phi_\ell^{\dag ab} (0,\bm{x}) | \underline{\rm n}; \bm{x}_1, \bm{x}_2 \rangle 
  \langle \underline{\rm n}; \bm{x}'_1, \bm{x}'_2 | \Phi_\ell^{bc}(0,\bm{x}) \left( \psi^\dag {\cal K}'_N T^c \chi \right) (\bm{x}) | \Omega \rangle 
\nonumber \\ 
&\hspace{5cm}
    = - V_{{\cal O}(N)} (\bm{x}_1, \bm{x}_2;\bm{\nabla}_1, \bm{\nabla}_2) \delta^{(3)} (\bm{x}_1-\bm{x}'_1) \delta^{(3)} (\bm{x}_2-\bm{x}'_2)\,,
\end{align} 
where, again, the contact terms $V_{{\cal O}(N)}(\bm{x}_1, \bm{x}_2;\bm{\nabla}_1, \bm{\nabla}_2)$ can be computed order by order in $1/m$.

\section{\boldmath $e^+e^- \to \chi_{QJ}(nP) + \gamma$}
We apply now the general framework to the computation of the cross sections for the $P$-wave quarkonium electromagnetic production processes  
\begin{equation}
e^+e^- \rightarrow \chi_{QJ}(nP)+\gamma\,.
\end{equation}
The NRQCD factorization formula up to relative order $v^2$ reads
\begin{align}
\sigma_{\chi_{QJ}+\gamma} = & ~~\sigma^{\rm em}_{Q \bar Q({}^3P_J^{[1]})} \,\langle \Omega | {\cal O}^{\chi_{QJ}}({}^3P_J^{[1]};\text{em}) | \Omega \rangle
                        + \sigma^{{\rm em}\,{\cal T}}_{Q \bar Q({}^3P_J^{[1]})} \, \langle \Omega | {\cal T}^{\chi_{QJ}}({}^3P_J^{[1]};\text{em}) | \Omega \rangle
\nonumber\\
                        &+ \sigma^{{\rm em}\,{\cal P}}_{Q \bar Q({}^3P_J^{[1]})} \,\langle \Omega | {\cal P}^{\chi_{QJ}}({}^3P_J^{[1]};\text{em}) | \Omega \rangle\,.
\end{align}
For a recent study of these cross sections in the framework of NRQCD we refer to~\cite{Brambilla:2017kgw},
where one may also find the expressions of the relevant four fermion operators.

Since the electromagnetic production matrix elements, $\langle \Omega| {\cal O}^{\cal Q}(N;\text{em})  |\Omega\rangle$,
are related to the electromagnetic decay matrix elements, $\langle {\cal Q}| {\cal O}^{\text{em}}(N)|{\cal Q}\rangle$, through
\begin{equation}
 \langle \Omega| {\cal O}^{\cal Q}(N;\text{em})  |\Omega\rangle = (2J+1) \langle {\cal Q}| {\cal O}^{\text{em}}(N)|{\cal Q}\rangle\,,
\end{equation}
where $J$ is the total angular momentum of the quarkonium $\cal Q$ and in the left-hand side a sum over all quarkonium polarizations is implicit, 
the computation of the production LDMEs is the same as the computation of the decay matrix elements for $P$-wave quarkonia~\cite{Brambilla:2001xy,Brambilla:2002nu,Brambilla:2020xod}.
Following~\cite{Brambilla:2020xod}, we find that after matching with pNRQCD the contact terms $ V_{{\cal O}(N)}(\bm{r},\bm{\nabla}_{\bm{r}})$ projecting on $P$-wave states read 
up to higher order corrections 
\begin{align}
 V_{{\cal O}^{\chi_{QJ}}({}^3P_J^{[1]};\text{em})}(\bm{r},\bm{\nabla}_{\bm{r}})=&
 N_c T_{1J}^{ij} \, \nabla_{\bm{r}}^i \, \left(1 + \frac{2}{3}  \frac{i {\cal E}_2}{m}   \right) \, \delta^{(3)} (\bm{r}) \nabla_{\bm{r}}^j\,,
  \\
  V_{{\cal T}^{\chi_{QJ}}({}^3P_J^{[1]};\text{em}) }(\bm{r},\bm{\nabla}_{\bm{r}}) =&
 N_c T_{1J}^{ij} \, \nabla_{\bm{r}}^i \, \frac{4}{3} \frac{{\cal E}_1}{m} \, \delta^{(3)} (\bm{r}) \nabla_{\bm{r}}^j\,,
  \\
  V_{{\cal P}^{\chi_{QJ}}({}^3P_J^{[1]};\text{em})}(\bm{r},\bm{\nabla}_{\bm{r}}) =&
 N_c T_{1J}^{ij} \, \nabla_{\bm{r}}^i \delta^{(3)} (\bm{r}) \, \left( - \bm{\nabla}_{\bm{r}}^2 - \frac{5}{3} {\cal E}_1 \right) \, \nabla_{\bm{r}}^j \,, 
\end{align}
where $T_{1J}^{ij}$ are spin projectors, 
\begin{align}
T_{10}^{ij} =& \frac{\sigma^i \otimes \sigma^j}{3} \,,\\
T_{11}^{ij} =& \frac{\epsilon_{kim}\epsilon_{kjn}\sigma^m \otimes \sigma^n}{2} \,,\\
T_{12}^{ij} =& \left(\frac{\delta_{im}\sigma^n + \delta_{in}\sigma^m}{2} -\frac{\delta_{mn}\sigma^i}{3}\right)
                \otimes \left(\frac{\delta_{jm}\sigma^n + \delta_{jn}\sigma^m}{2} - \frac{\delta_{mn}\sigma^j}{3}\right)\,,
\end{align}                
and ${\cal E}_n$ are correlators of two chromoelectric fields $E^{a,i}$ located at $\bm{0}$:
\begin{equation}
{\cal E}_n = \frac{1}{2N_c} \int_0^\infty dt \, t^n \langle \Omega | g E^{a,i} (t) \Phi^{ab}(0;t) g E^{b,i} (0) | \Omega\rangle\,,
\end{equation}
$\Phi^{ab}(0;t)$ is a Wilson line in the adjoint representation connecting $(0,\bm{0})$ with $(t,\bm{0})$.

It follows from \eqref{LDMEpNRQCD} that the pNRQCD factorization formulas for $P$-wave quarkonium electromagnetic production read
\begin{align}
  \langle \Omega | {\cal O}^{\chi_{QJ}}({}^3P_J^{[1]};\text{em}) | \Omega \rangle =& (2J+1) \frac{3 N_c}{2 \pi} |R'(\bm{0})|^2 
\left[ 1 + \frac{2}{3} \frac{i{\cal E}_2}{m} + O\left(v^2\right) \right]\,,
  \\
  \langle \Omega | {\cal T}^{\chi_{QJ}}({}^3P_J^{[1]};\text{em}) | \Omega \rangle =&  (2J+1) \frac{3 N_c}{2 \pi} |R'(\bm{0})|^2 \frac{4}{3} \frac{{\cal E}_1}{m}
                                                                                     \left[ 1 +  O\left(v\right) \right]\,,
  \\
\langle \Omega | {\cal P}^{\chi_{QJ}}({}^3P_J^{[1]};\text{em}) | \Omega \rangle =& (2J+1) \frac{3 N_c}{2 \pi} | R'(\bm{0})|^2 
\left[ m \varepsilon - \frac{2}{3} {\cal E}_1 + O\left(v^3\right) \right]\,;
\end{align}
$R'(\bm{0})$ is the derivative of the $P$-wave radial wavefunction at the origin, and $\varepsilon$ is the binding energy.

\subsection{Correlators}
Both $R'(\bm{0})$ and $\varepsilon$ can be computed out of the eigenfunctions and eigenvalues of the pNRQCD Hamiltonian $h_0(\bm{x}_1, \bm{x}_2; \bm{\nabla}_1, \bm{\nabla}_2)$.
The potential should ideally be determined from lattice QCD.
This becomes unpractical, however, if $R'(\bm{0})$ needs to be known beyond leading order in $v$,
because not all relativistic corrections to the quarkonium potential have been computed in lattice QCD.
In~\cite{Brambilla:2020xod}, $R'(\bm{0})$ and $\varepsilon$ were determined using several potential models.
The chromoelectric field correlators were instead fitted on the decay widths for $\chi_{c0}(1P) \to \gamma\gamma$, $\chi_{c2}(1P) \to \gamma\gamma$,
taken from the PDG~\cite{ParticleDataGroup:2020ssz},
and on $\sigma(e^+e^- \rightarrow \chi_{c1}(1P) + \gamma) = 17.3^{+4.2}_{-3.9}\pm 1.7$~fb at $\sqrt{s}=$10.6~GeV, taken from from Belle~\cite{Belle:2018jqa},
obtaining 
\begin{align}
{\cal E}_1  =& -0.20 {}^{+0.14}_{-0.14} \pm 0.90 \textrm{ GeV}^2 \label{E1final}\,,\\
i{\cal E}_2 =& 0.77 {}^{+0.98}_{-0.86} \pm 0.85 \textrm{ GeV}  \label{E2final}\,,
\end{align}
where the first uncertainty accounts for the model dependence.

\subsection{Phenomenology}
\begin{figure}[ht]
\begin{center}
\epsfxsize=9truecm \epsfbox{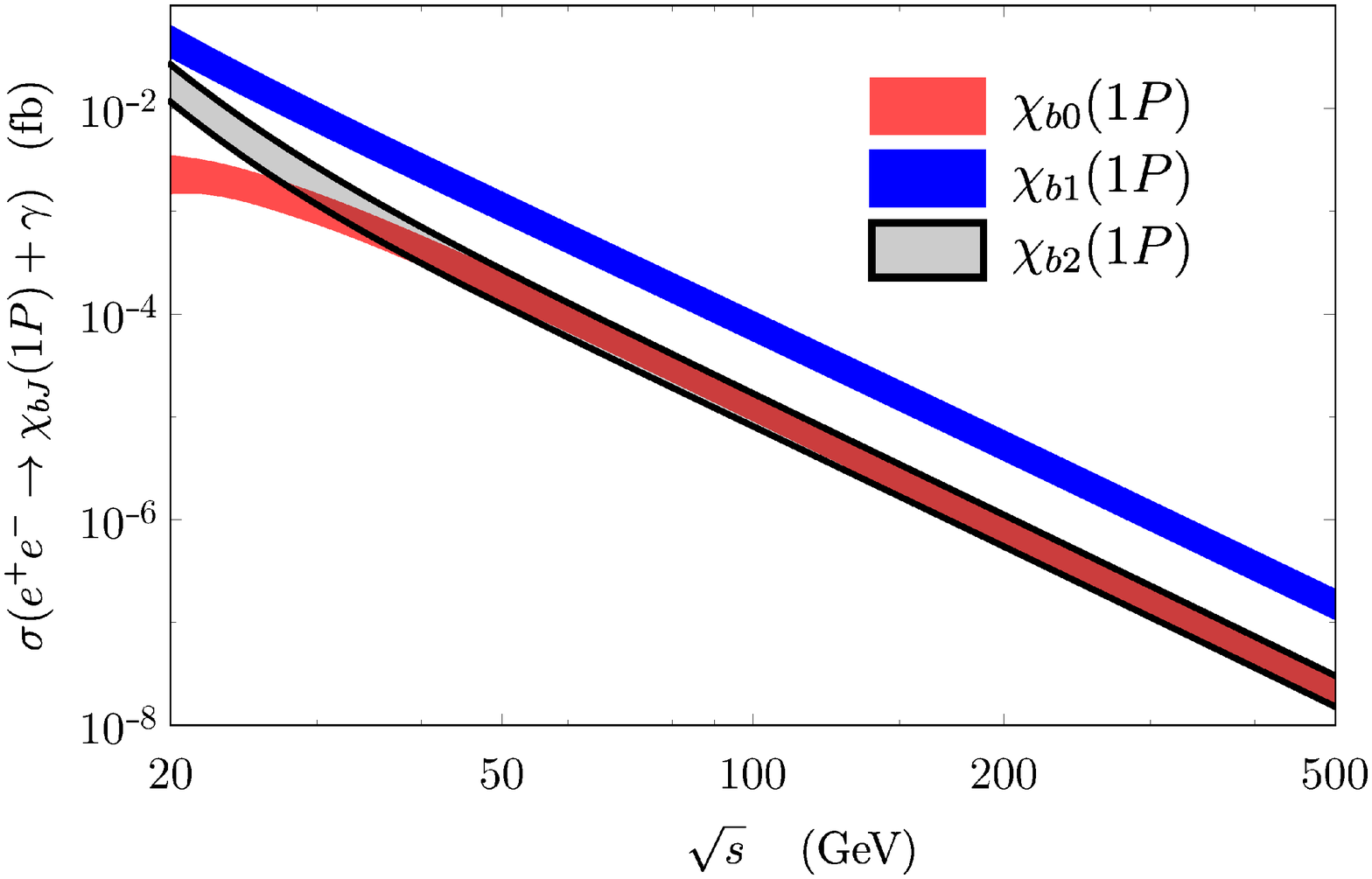}
\end{center}
\caption{
  Predicted $\sigma(e^+ e^- \to \chi_{bJ}(1P) + \gamma)$ for $J=0$, $1$, $2$.
  From~\cite{Brambilla:2020xod}.
  \label{fig:cs_chib1P}
}
\end{figure}

\begin{figure}[ht]
\begin{center}
\epsfxsize=9truecm \epsfbox{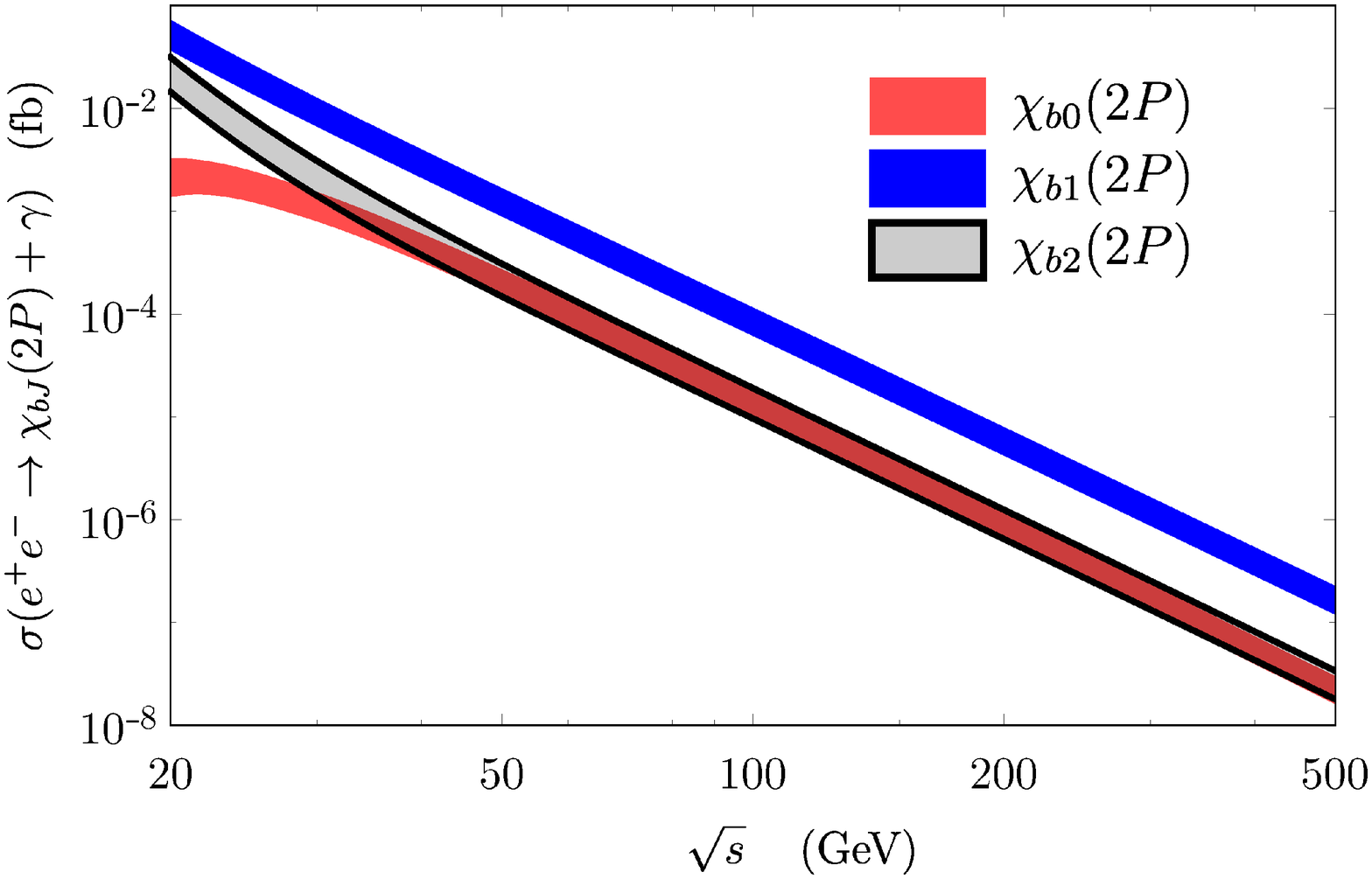}
\end{center}
\caption{
  Predicted $\sigma(e^+ e^- \to \chi_{bJ}(2P) + \gamma)$ for $J=0$, $1$, $2$.
  From~\cite{Brambilla:2020xod}.
  \label{fig:cs_chib2P}}
\end{figure}

\begin{figure}[ht]
\begin{center}
\epsfxsize=9truecm \epsfbox{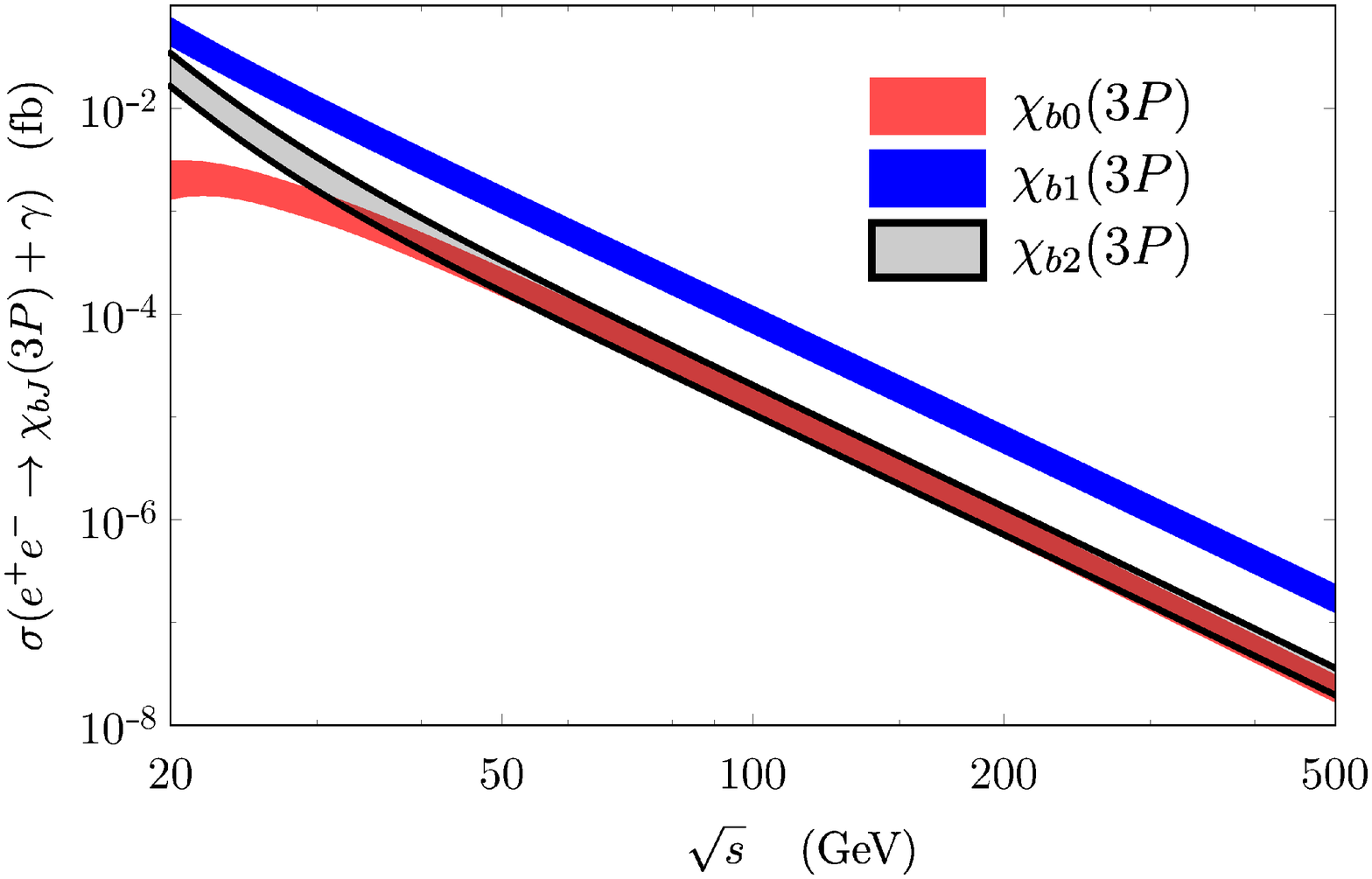}
\end{center}
\caption{
  Predicted $\sigma(e^+ e^- \to \chi_{bJ}(3P) + \gamma)$ for $J=0$, $1$, $2$. 
  From~\cite{Brambilla:2020xod}.
  \label{fig:cs_chib3P}}
\end{figure}

The correlators are universal: they do not depend neither on the flavor of the heavy quark nor on the quarkonium state.
It is precisely the universal nature of the correlators
that allows us to use them to compute cross sections (and decay widths) for quarkonia with different principal quantum number and bottomonia.
Summarizing our results, in the charmonium sector we get at the Belle center of mass energy $\sqrt{s}=$10.6~GeV
the following $P$-wave charmonium cross sections~\cite{Brambilla:2020xod}:
\begin{align}
\sigma(e^+ e^- \to \chi_{c0}(1P) + \gamma) =& 1.84 {}^{+0.25}_{-0.26} \pm 0.76 \textrm{~fb}\,,\\
\sigma(e^+ e^- \to \chi_{c1}(1P) + \gamma) =& 16.4 {}^{+0.2}_{-0.2}  \pm 6.4 \textrm{~fb} \,,\\
\sigma(e^+ e^- \to \chi_{c2}(1P) + \gamma) =& 3.75 {}^{+0.67}_{-0.56} \pm 2.16 \textrm{~fb}\,.
\end{align}
The results for the corresponding cross sections in the bottomonium sector are shown for a wide range of center of mass energies
in figures~\ref{fig:cs_chib1P}, \ref{fig:cs_chib2P} and~\ref{fig:cs_chib3P}.

\section{\boldmath $pp \to \chi_{QJ}(nP)  + X$}
\begin{figure}[ht]
\begin{center}
\includegraphics[width=0.4\textwidth]{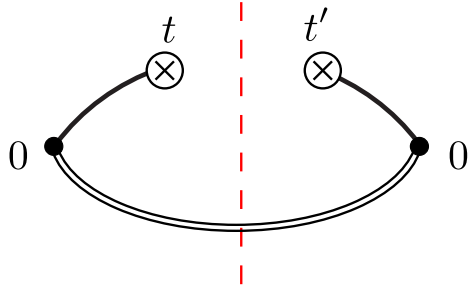} 
\caption{Graphical representation of the chromoelectric fields and Wilson lines in equation~\eqref{eq:ftensor}: 
a circle with a cross represents an insertion of a chromoelectric field at the given time, the filled circle represents the spacetime origin, 
solid lines stand for temporal Wilson lines, double lines for gauge-completion Wilson lines in the $\ell$ direction, and the dashed line is the cut.
\label{fig:cut}}
\end{center}
\end{figure}

We consider the quarkonium inclusive hadroproduction processes
\begin{equation}
  pp \rightarrow h_{Q}(nP)+X \qquad \text{and} \qquad pp \rightarrow \chi_{QJ}(nP)+X \,.
\end{equation}
The NRQCD factorization formulas at leading order in $v$ read~\cite{Bodwin:1994jh}
\begin{align}
  \sigma_{h_Q+X} &= \sigma_{Q \bar Q({}^1P_1^{[1]})} \langle \Omega | {\cal O}^{h_Q}({}^1P_1^{[1]}) | \Omega \rangle
                   + \sigma_{Q \bar Q({}^1S_0^{[8]})} \langle \Omega | {\cal O}^{h_Q}({}^1S_0^{[8]}) | \Omega \rangle\,,
\label{sigmahQX}                         \\
  \sigma_{\chi_{QJ}+X} &= \sigma_{Q \bar Q({}^3P_J^{[1]})} \langle \Omega | {\cal O}^{\chi_{QJ}}({}^3P_J^{[1]}) | \Omega \rangle
                         + \sigma_{Q \bar Q({}^3S_1^{[8]})} \langle \Omega | {\cal O}^{\chi_{QJ}}({}^3S_1^{[8]}) | \Omega \rangle\,.
\label{sigmachiQX}
\end{align}
Following~\cite{Brambilla:2020ojz,Brambilla:2021abf},
we find that after matching with pNRQCD the contact terms $ V_{{\cal O}(N)}(\bm{r},\bm{\nabla}_{\bm{r}})$ projecting on $P$-wave states read 
up to higher order corrections 
\begin{align}
V_{{\cal O} ({}^1P_1^{[1]})}(\bm{r},\bm{\nabla}_{\bm{r}}) =& N_c \nabla_{\bm{r}}^i \delta^{(3)} (\bm{r}) \nabla_{\bm{r}}^i \,,\\
V_{{\cal O} ({}^1S_0^{[8]})}(\bm{r},\bm{\nabla}_{\bm{r}}) =& N_c \nabla_{\bm{r}}^i \delta^{(3)} (\bm{r}) \nabla_{\bm{r}}^j \frac{ {\cal E}^{ij} }{N_c^2 m^2}\,,\\
V_{{\cal O} ({}^3P_J^{[1]})}(\bm{r},\bm{\nabla}_{\bm{r}}) =& T_{1J}^{ij} N_c \nabla_{\bm{r}}^i \delta^{(3)} (\bm{r}) \nabla_{\bm{r}}^j \,,\\ 
V_{{\cal O} ({}^3S_1^{[8]})}(\bm{r},\bm{\nabla}_{\bm{r}}) =& \sigma^k \otimes \sigma^k N_c \nabla_{\bm{r}}^i \delta^{(3)} (\bm{r}) \nabla_{\bm{r}}^j \frac{{\cal E}^{ij}}{N_c^2 m^2}\,,
\end{align} 
where the tensor ${\cal E}^{ij}$ is defined by 
\begin{align}
\label{eq:ftensor} 
{\cal E}^{ij} &= 
\int_0^\infty dt\, t \; \int_0^\infty dt'\, t'\; \langle \Omega |\Phi_\ell^{\dag ab} \Phi^{\dag ad} (0;t) g E^{d,i}(t) g E^{e,j}(t') \Phi^{ec} (0;t') \Phi_\ell^{bc} | \Omega \rangle\,.
\end{align} 
The direction component $\ell^0$ may be chosen in such a way that the fields in $g E^{e,j}(t') \Phi^{ec} (0;t') \Phi_\ell^{bc}$ are time ordered (${\cal T}$)
and those in $\Phi_\ell^{\dag ab} \Phi^{\dag ad} (0;t) g E^{d,i}(t)$ are anti-time ordered ($\bar{\cal T}$).
Hence the correlator ${\cal E}^{ij}$ may be interpreted as the cut diagram shown in figure~\ref{fig:cut}.
For polarization-summed cross sections or for production of scalar states only the isotropic part of ${\cal E}^{ij}$ is relevant.
This is  the dimensionless chromoelectric correlator ${\cal E}$:
\begin{equation}
{\cal E} = \frac{3}{N_c} \int_0^\infty dt\, t \; \int_0^\infty dt'\, t' \;
\langle \Omega | \Phi_\ell^{\dag ab} \Phi^{\dag ad} (0;t) g E^{d,i}(t) g E^{e,i}(t') \Phi^{ec} (0;t') \Phi_\ell^{bc} | \Omega \rangle \,.
\end{equation}

It follows from \eqref{LDMEpNRQCD} that the pNRQCD factorization formulas for $P$-wave quarkonium hadroproduction read at leading order in $v$ (and $1/N_c$ for octet LDMEs)
\begin{align}
\langle \Omega | {\cal O}^{h_Q}({}^1P_1^{[1]}) | \Omega \rangle =& 3 \, \frac{3 N_c}{2\pi} |R'(\bm{0})|^2\,,\\
\langle \Omega  |{\cal O}^{h_Q} ({}^1S_0^{[8]}) | \Omega \rangle =& 3 \, \frac{3 N_c}{2 \pi} |R'(\bm{0})|^2 \frac{\cal E}{9 N_c m^2} \,,\\
\langle \Omega | {\cal O}^{\chi_{QJ}}({}^3P_J^{[1]}) | \Omega \rangle =& (2 J+1) \, \frac{3 N_c}{2\pi} |R'(\bm{0})|^2\,,\\ 
\langle \Omega | {\cal O}^{\chi_{QJ}}({}^3S_1^{[8]}) | \Omega \rangle =& (2 J+1) \, \frac{3 N_c}{2 \pi} |R'(\bm{0})|^2 \frac{\cal E}{9 N_c m^2}\,.
\end{align}
The LDMEs have to be understood as polarization summed.
The above expressions imply (at leading order in $v$) the universality of the ratios
\begin{align} 
\frac{m^2\langle \Omega | {\cal O}^{\chi_{QJ}}({}^3S_1^{[8]}) | \Omega \rangle} {\langle \Omega | {\cal O}^{\chi_{QJ}}({}^3P_J^{[1]}) | \Omega \rangle} 
= \frac{m^2\langle \Omega | {\cal O}^{h_Q}({}^1S_0^{[8]}) | \Omega \rangle}{\langle \Omega | {\cal O}^{h_Q}({}^1P_1^{[1]}) | \Omega \rangle} 
= \frac{\cal E}{9 N_c} \,.
\end{align}

\subsection{NRQCD factorization}
For the pNRQCD expressions of the LDMEs to be consistent with perturbative QCD, they must reproduce the same infrared divergences.
At two loop accuracy and at lowest order in the relative momentum $q$ of the heavy quark and antiquark, the infrared diverges in the NRQCD LDMEs can be cast in the factor 
\begin{align} 
{\cal I}_2 (p,q) = & 
\sum_N \int_0^\infty d \lambda' \, \lambda' 
\langle \Omega | \bar{\cal T} \left\{ \Phi_\ell^{\dag c'b} \Phi_p^{\dag a'c'}(\lambda') [p^\mu q^\nu F_{\nu \mu}^{a'} (\lambda'p) ] \right\} | N \rangle 
\nonumber\\ 
& \times 
\int_0^\infty d \lambda \, \lambda \langle N | {\cal T} \left\{ \Phi_\ell^{bc} [ p^\mu q^\nu F_{\nu \mu}^{a} (\lambda p)] \Phi_p^{ac} (\lambda) \right\} | \Omega\rangle\,,
\end{align}
where the sum in $N$ goes over all possible intermediate states, $p$ is half the center of mass momentum of the heavy quark-antiquark pair and 
$\Phi_p (\lambda)$ is an adjoint Wilson line along $p$ connecting $0$ with $\lambda p$~\cite{Nayak:2005rw,Nayak:2005rt,Nayak:2006fm}.

Since in ${\cal I}_2(p,q)$ a momentum $q$ comes from each side of the cut, the factor ${\cal I}_2 (p,q)$ contributes to the production of color singlet $P$-wave states.
In the rest frame of the heavy quark-antiquark pair, $\bm{p} = 0$, $q^0 = 0$, $\Phi_p (\lambda) = \Phi(0;t)$ with $t = \sqrt{p^2} \lambda$,
$p^\mu q^\nu F_{\nu \mu}^{a} (\lambda p) = -\sqrt{p^2} q^i E^{a,i} (t)$ so that ${\cal I}_2(p,q)$ can be written as ${\cal E}^{ij} \, q^i q^j/p^2$.
Since this expression is proportional to the contact terms $V_{{\cal O}(^1S_0^{[8]})}$ and $V_{{\cal O}(^3S_1^{[8]})}$ in momentum space, 
the pNRQCD expressions for the color octet LDMEs reproduce the infrared divergences of the NRQCD infrared factor.
Moreover, the one-loop running of $\cal E$ is $\displaystyle \frac{d}{d \log \Lambda} {\cal E} (\Lambda) = 12 C_F \frac{\alpha_{\text{s}}}{\pi}$, 
which implies $\displaystyle \frac{d}{d \log \Lambda} \langle {\cal O}^{\chi_{QJ}} ({}^3S_1^{[8]}) \rangle
= \frac{4 C_F \alpha_{\text{s}}}{3 N_c \pi m^2} \langle {\cal O}^{\chi_{QJ}} ({}^3P_J^{[1]}) \rangle$ with $C_F=(N_c^2-1)/(2N_c)$.
This agrees with the one-loop evolution equation derived in perturbative NRQCD~\cite{Bodwin:1994jh}.

\subsection{Correlator}
Following~\cite{Brambilla:2021abf}, the correlator $\cal E$ can be fitted to the ratio $r_{21}\! = \! (d \sigma_{\chi_{c2}(1P)}/dp_T)/(d \sigma_{\chi_{c1}(1P)}/dp_T)$,
which does not depend (at leading order in $v$) on the wavefunction.
In order to compare to measurements, we compute $r_{21}$ multiplied with $B_{\chi_{c2}(1P)}/ B_{\chi_{c1}(1P)}$,
where $B_{\chi_{cJ}(1P)} = {\rm Br} (\chi_{cJ(1P)} \to J/\psi \gamma) \, {\rm Br} (J/\psi \to \mu^+ \mu^-)$, and Br stands for the branching ratio.  
We take $B_{\chi_{cJ}(1P)}$ from the PDG~\cite{ParticleDataGroup:2020ssz}. 

\begin{figure}[ht] 
\begin{center}
\includegraphics[width=0.6\textwidth]{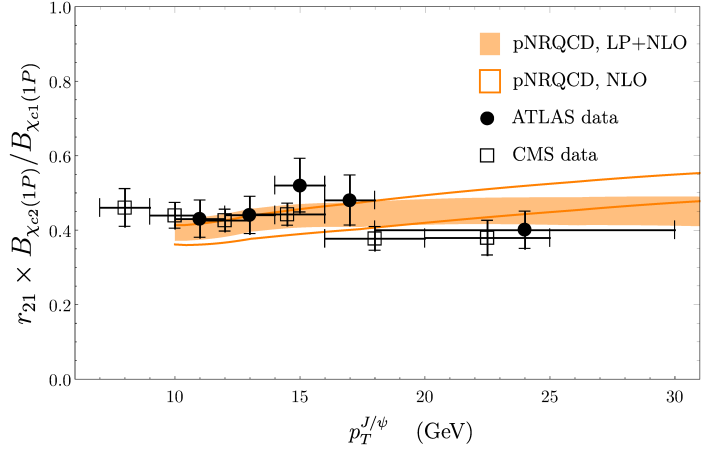}
\caption{
  The ratio of the $\chi_{c2}(1P)$ and $\chi_{c1}(1P)$ differential cross sections times $B_{\chi_{c2}(1P)}/ B_{\chi_{c1}(1P)}$
  at the LHC center of mass energy $\sqrt{s}=7$~TeV and in the rapidity range $|y|<0.75$,
  with fitted ${\cal E}$, compared to CMS~\cite{CMS:2012qwg} and ATLAS~\cite{ATLAS:2014ala} data. From~\cite{Brambilla:2021abf}.
  \label{fig:chicratio}} 
\end{center}
\end{figure}

By performing a fit to the measured values of $r_{21} \, B_{\chi_{c2}(1P)}/ B_{\chi_{c1}(1P)}$ by CMS~\cite{CMS:2012qwg} and ATLAS~\cite{ATLAS:2014ala}, 
we obtain, from fixed-order next-to-leading order (NLO) calculations of the short-distance coefficients,
\begin{equation} 
\label{eq:corrE_value} 
{\cal E}|_{\rm NLO}(\Lambda=1.5\textrm{~GeV}) = 1.17 \pm 0.05\,.
\end{equation} 
Alternatively, we can use short-distance coefficients where logarithms in the ratio of the transverse momentum, $p_T$, over the charm mass, $m_c$,
have been resummed at leading logarithmic accuracy at leading power (LP) in the expansion in powers of $m_c/p_T$~\cite{Bodwin:2014gia,Bodwin:2015iua}.
Although these coefficients contain more information, large LP contributions are generated at order $\alpha_\text{s}^2$, which is incomplete.
By using LP+NLO expressions for the short-distance coefficients, we obtain
\begin{equation}
\label{eq:corrE_value2}
{\cal E}|_{\rm LP+NLO}(\Lambda=1.5\textrm{~GeV}) = 4.48 \pm 0.14\,.
\end{equation}
The difference between the values of ${\cal E}$ in equations~\eqref{eq:corrE_value} and~\eqref{eq:corrE_value2} reflects 
the difference between fixed-order NLO and LP+NLO calculations of the short-distance coefficients.
We show our result for $r_{21}$ compared to ATLAS and CMS data in figure~\ref{fig:chicratio}.

The combination of the two determinations of ${\cal E}$ gives 
\begin{equation}
\label{eq:corrE_value3}
{\cal E}(\Lambda=1.5\textrm{~GeV}) = 2.8 \pm 1.7\,, 
\end{equation}
where the central value is the average of the central values of the determinations in~\eqref{eq:corrE_value} and~\eqref{eq:corrE_value2},
and the error is such to encompass both determinations.
We will use the combination \eqref{eq:corrE_value3} when computing bottomonium cross sections.

\subsection{Phenomenology}
The correlator ${\cal E}$ is universal: it does not depend neither on the flavor of the heavy quark nor on the quarkonium state.
As in the electromagnetic production case, the universal nature of the correlator allows us to use it to compute cross sections for quarkonia with different
principal quantum number and for bottomonia (once accounted for the running) without having to fit new octet LDMEs.

The following hadroproduction results are based on the formulas \eqref{sigmahQX} and \eqref{sigmachiQX}, which are valid at leading order in $v$.
Hence we just need the $P$-wave quarkonium wavefunction at the origin at leading order in $v$.
In the charmonium case, it may be extracted from $\Gamma(\chi_{c0,2}(1P) \to \gamma \gamma)$ data.
In the bottomonium case, we take it from the set of potential models considered in~\cite{Brambilla:2020xod}. 

\begin{figure}[ht] 
\begin{center}
\includegraphics[width=0.6\textwidth]{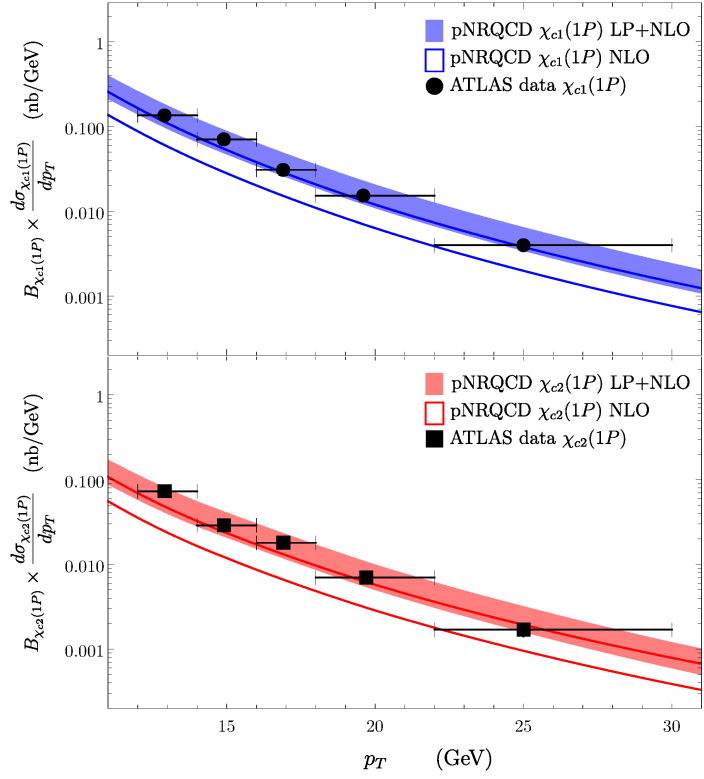}
\caption{
  Differential production cross sections of the $\chi_{c1}(1P)$ and $\chi_{c2}(1P)$ at the LHC center of mass energy $\sqrt{s}=7$~TeV
  and in the rapidity range $|y|<0.75$ compared with ATLAS data~\cite{ATLAS:2014ala}. From~\cite{Brambilla:2021abf}.
  \label{fig:chicrate}} 
\end{center}
\end{figure}

\begin{figure}[ht] 
\begin{center}
\includegraphics[width=0.6\textwidth]{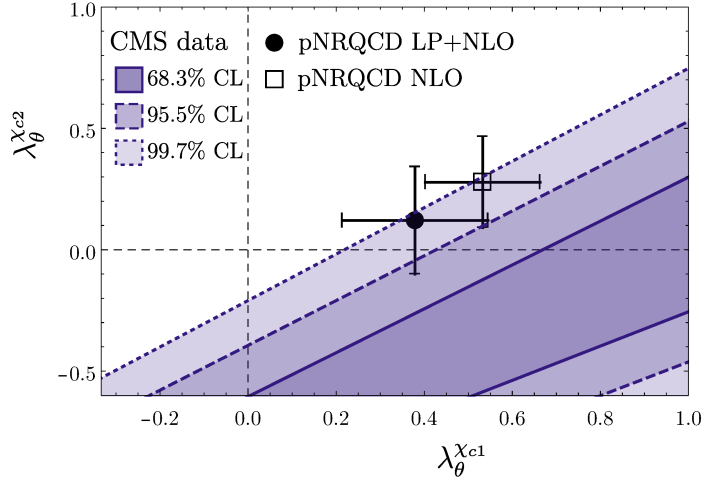} 
\caption{
  The polarization parameters $\lambda_\theta^{\chi_{c1}}$ and $\lambda_\theta^{\chi_{c2}}$ at the LHC center of mass energy $\sqrt{s}=7$~TeV and in the rapidity range $|y|<0.75$,
  averaged over the $J/\psi$ transverse momentum range 8~GeV$<p_T^{J/\psi}<$30~GeV, compared with constraints from CMS~\cite{CMS:2019jas}. From~\cite{Brambilla:2021abf}.
  \label{fig:chicpol}}  
\end{center}
\end{figure}

For $P$-wave charmonium production, we show in figure~\ref{fig:chicrate} the differential production cross sections $\sigma(pp \to \chi_{cJ}(1P) + X)$ for $J=1$, $2$.
We may also consider polarized cross sections.
For polarized cross sections, the non-isotropic part of ${\cal E}^{ij}$ can in principle contribute to the color octet matrix elements,
and, if such contribution does not vanish, the color octet matrix elements acquire a dependence on the direction of the gauge-completion Wilson lines.  
For the universality of the NRQCD LDMEs to be valid also for the case of polarized cross sections,
such non-isotropic contributions need to vanish in the NRQCD matrix elements.
This has not been proved, but often assumed in the literature, leading to the equalities 
\begin{align}
\langle \Omega | {\cal O}^{h_{Q}}({}^1S_0^{[8]}) | \Omega \rangle
=& 3 \, \langle \Omega | \chi^\dag T^a \psi \Phi_\ell^{\dag ab} (0) {\cal P}_{{h_{Q}}(\lambda, \bm{P}=\bm{0})} \Phi_\ell^{bc} (0) \psi^\dag T^c \chi | \Omega \rangle\,,
\\
\langle \Omega | {\cal O}^{\chi_{QJ}}({}^3S_1^{[8]}) | \Omega \rangle 
=& (2J+1) \, \langle \Omega | \chi^\dag \sigma^i T^a \psi \Phi_\ell^{\dag ab} (0) {\cal P}_{{\chi_{QJ}}(\lambda, \bm{P}=\bm{0})} \Phi_\ell^{bc} (0) \psi^\dag \sigma^i T^c \chi | \Omega \rangle\,.
\end{align} 
Under the assumption of universality of the polarized color octet LDMEs, we can compute the polarization parameters
\begin{equation}
\lambda_\theta^{\chi_{cJ}} = \frac{1 - 3 \xi_{\chi_{cJ}}}{1 + \xi_{\chi_{cJ}}}\,,
\end{equation}
where $\xi_{\chi_{cJ}}$ is the fraction of $J/\psi$ produced with longitudinal polarization from decays of $\chi_{cJ}$.
The spin quantization axis of the $J/\psi$ is defined in the hadron helicity frame.
The result is shown in figure~\ref{fig:chicpol}.

\begin{figure}[ht] 
\begin{center}
\includegraphics[width=0.6\textwidth]{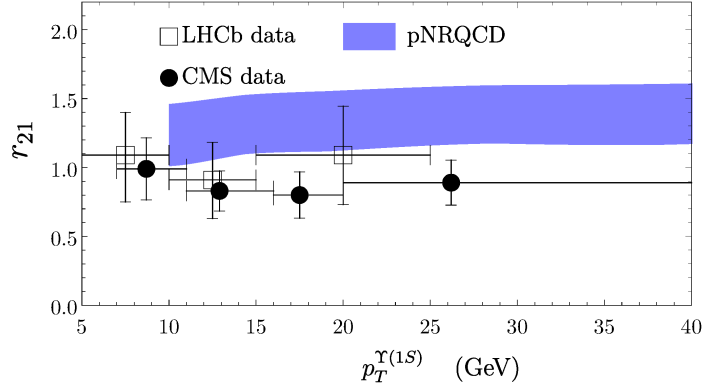}
\caption{
  Ratio of $\chi_{b2}(1P)$ and $\chi_{b1}(1P)$ differential cross sections at the LHC center of mass energy $\sqrt{s}=7$~TeV
  and in the rapidity range $2 < y < 4.5$ compared with LHCb~\cite{LHCb:2014nug} and CMS~\cite{CMS:2014bsd} data. From~\cite{Brambilla:2021abf}.
  \label{fig:chibratio}} 
\end{center}
\end{figure}

\begin{figure}[ht] 
\begin{center}
\includegraphics[width=0.6\textwidth]{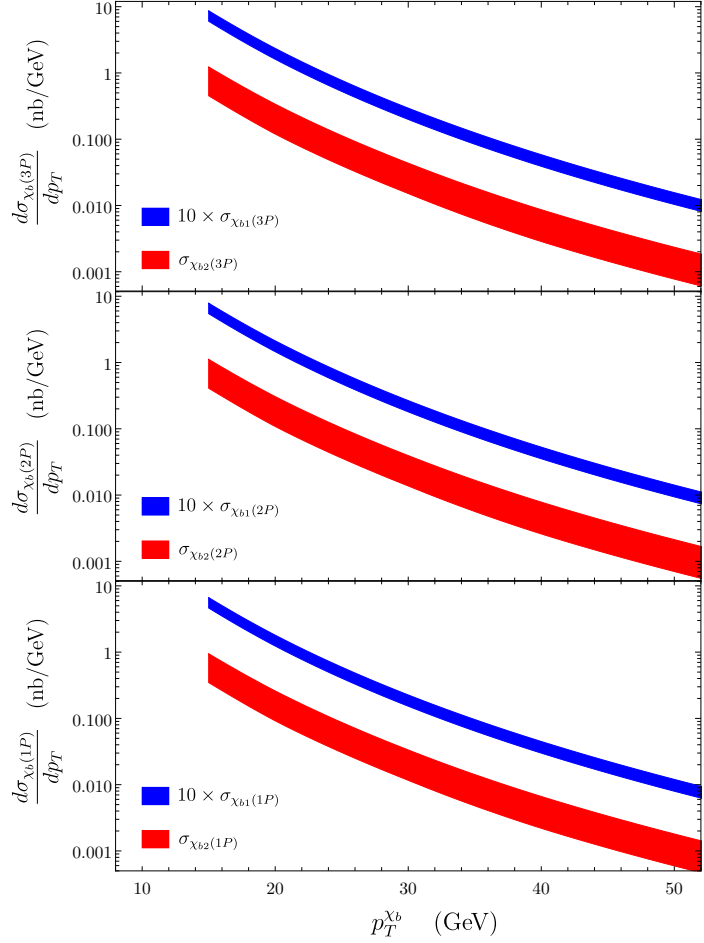}
\caption{Production cross sections of the $\chi_{b1} (nP)$ and $\chi_{b2} (nP)$ ($n=1$, 2, and 3) at the
  LHC center of mass energy $\sqrt{s}=7$~TeV in the rapidity range $2 < y < 4.5$. From~\cite{Brambilla:2021abf}.
    \label{fig:chibabs}} 
\end{center}
\end{figure}

\begin{figure}[ht] 
\begin{center}
  \includegraphics[width=0.6\textwidth]{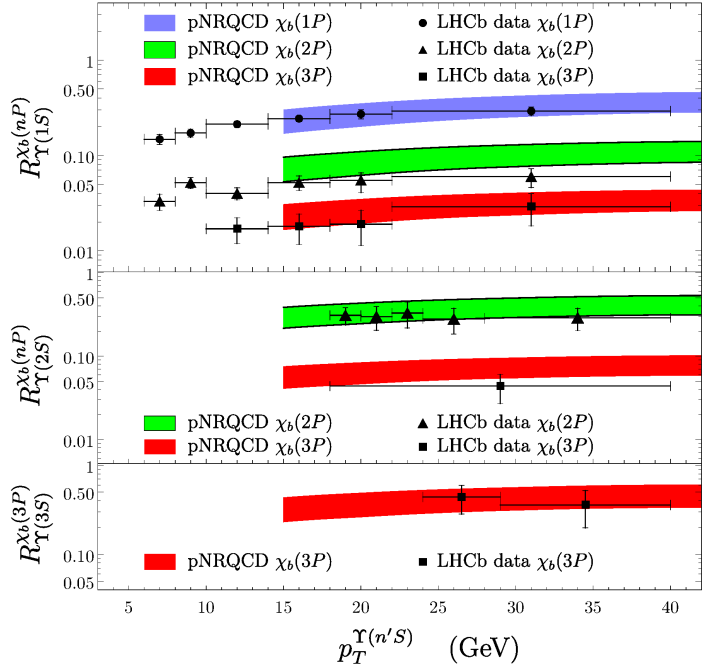}
\caption{
  Feeddown fractions $R_{\Upsilon(n'S)}^{\chi_b(nP)}$ at the LHC center of mass energy $\sqrt{s}=7$~TeV and in the rapidity range $2 < y < 4.5$ 
  compared with LHCb data~\cite{LHCb:2014ngh}. From~\cite{Brambilla:2021abf}.
    \label{fig:chibrate}} 
\end{center}
\end{figure}

For $P$-wave bottomonium, we show in figure~\ref{fig:chibratio} the differential production cross section ratio $(d \sigma_{\chi_{b2}(1P)}/dp_T)/(d \sigma_{\chi_{b1}(1P)}/dp_T)$,
which does not depend at leading order in $v$ and $1/N_c$ on the bottomonium wavefunction and, therefore, is expected to be the same for all $nP$ bottomonium states.
We compare with LHCb and CMS data.
The ratio does depend instead on $\cal E$ (at the scale of the $b$ mass).
Since $\cal E$ has been determined on charmonium data (we use here the value given in equation~\eqref{eq:corrE_value3}),
this is a test of the universality of the pNRQCD factorization.
In figure~\ref{fig:chibabs}, we predict the inclusive hadroproduction differential cross sections $pp \to \chi_{bJ}(nP) + X$ for $J=1$, $2$ and $n=1$, $2$, $3$,
which have not been measured yet.
Finally, in figure~\ref{fig:chibrate} we compare with LHCb data the feeddown fractions 
$\displaystyle R_{\Upsilon(n'S)}^{\chi_{b}(nP)} =
\frac{\sum_{J=1,2} {\rm Br}({\chi_{bJ} (nP) \to \Upsilon(n'S)+\gamma})  \, \sigma_{\chi_{bJ}(nP)} } { \sigma_{\Upsilon(n'S)}}$.
Note that for the feeddown fractions model dependence enters not only in the $\chi_{bJ}$ wavefunctions but also to some extent in the
determination of the $S$-wave bottomonium cross sections and in some branching ratios.

\section{Outlook}
We have reviewed how strongly coupled pNRQCD can be used to factorize certain NRQCD production matrix elements into universal correlators and
a factor that depends on the behaviour of the wavefunction at the origin.
The result is similar to an analogous factorization valid for NRQCD decay matrix elements, although the correlators in the hadroproduction case are different.
From a theoretical perspective,
the pNRQCD factorization of the NRQCD production matrix elements in the case of hadroproduction may shed some light on their conjectured universality. 
From the phenomenology viewpoint, pNRQCD reduces the number of unknown parameters in the expressions of the quarkonium production cross sections;
more specifically, the novelty of the pNRQCD approach resides in the treatment of the color octet long distance matrix elements.
Hence the pNRQCD expressions have more predictive power than the corresponding NRQCD ones.
In particular, correlators determined with charmonium data may be used to compute color octet matrix elements and hence observables in the bottomonium sector.
A clean example of this is the determination of the ratio of the $\chi_{b2}(1P)$ and $\chi_{b1}(1P)$ differential cross sections for hadroproduction 
shown in figure~\ref{fig:chibratio}.

Ideally, correlators of gluon fields should be determined by computations in lattice QCD.
There exist some specific lattice studies for few of the correlators entering quarkonium electromagnetic production, but none for those entering quarkonium hadroproduction.
The major uncertainties at the moment come, however, from the poor knowledge of the quarkonium wavefunctions.
At leading order in the velocity, $P$-wave charmonium wavefunctions at the origin may be extracted from two photon decay data, 
but neither the $\chi_{b0}$ and $\chi_{b2}$ decay widths into two photons nor their total decay widths have been measured yet.
Hence $P$-wave bottomonium wavefunctions rely on potential models already at leading order.
Beyond leading order, also charmonium wavefunctions are presently model dependent.
Recent progress towards a more rigorous determination of the behaviour of the quarkonium wavefunctions at the origin has been made in~\cite{Chung:2020zqc,Chung:2021efj}.

The pNRQCD factorization has been applied so far to $P$-wave quarkonium production only.
An obvious extension would be the treatment of $S$-wave quarkonium production, including some relevant electroweak channel (like in~\cite{Brambilla:2019fmu}).
For $S$-wave quarkonium production a large amount of data is available and yet a completely satisfactory description in the framework of NRQCD is missing,
an example being hadroproduction of the $\eta_c$.
Since for $S$-wave production color octet matrix elements show up beyond leading order in the velocity expansion,
pNRQCD factorization starts playing a crucial role at relative order $v^2$ or higher.
Although phenomenological applications may be then limited by the poor knowledge of the quarkonium wavefunctions beyond leading order,
as we discussed above, still pNRQCD can provide stringent constraints on the $S$-wave color octet matrix elements.

\acknowledgments
I thank Nora Brambilla, Hee Sok Chung and Daniel M\"uller for collaboration on the work presented here.
This work has been funded by the DFG Project-ID 196253076 - TRR 110.


\begin{thebibliography}{99}

\bibitem{Brambilla:2004jw}
N.~Brambilla, A.~Pineda, J.~Soto and A.~Vairo,
{\it Effective field theories for heavy quarkonium},
Rev. Mod. Phys. \textbf{77} (2005) 1423
[arXiv:hep-ph/0410047 [hep-ph]].

\bibitem{QuarkoniumWorkingGroup:2004kpm}
N.~Brambilla \text{et al.} [Quarkonium Working Group],
{\it Heavy quarkonium physics},
[arXiv:hep-ph/0412158 [hep-ph]].

\bibitem{Brambilla:2010cs}
N.~Brambilla, S.~Eidelman, B.~K.~Heltsley, R.~Vogt, G.~T.~Bodwin, E.~Eichten, A.~D.~Frawley, A.~B.~Meyer, R.~E.~Mitchell and V.~Papadimitriou, \text{et al.}
{\it Heavy quarkonium: progress, puzzles, and opportunities},
Eur. Phys. J. C \textbf{71} (2011) 1534
[arXiv:1010.5827 [hep-ph]].

\bibitem{Bodwin:2013nua}
G.~T.~Bodwin, E.~Braaten, E.~Eichten, S.~L.~Olsen, T.~K.~Pedlar and J.~Russ,
{\it Quarkonium at the frontiers of high energy physics: a Snowmass white paper},
[arXiv:1307.7425 [hep-ph]].

\bibitem{Brambilla:2020xod}
N.~Brambilla, H.~S.~Chung, D.~M\"uller and A.~Vairo,
{\it Decay and electromagnetic production of strongly coupled quarkonia in pNRQCD},
JHEP \textbf{04} (2020) 095
[arXiv:2002.07462 [hep-ph]].

\bibitem{Brambilla:2020ojz}
N.~Brambilla, H.~S.~Chung and A.~Vairo,
{\it Inclusive hadroproduction of $P$-wave heavy quarkonia in potential nonrelativistic QCD},
Phys. Rev. Lett. \textbf{126} (2021) no.8, 082003
[arXiv:2007.07613 [hep-ph]].

\bibitem{Brambilla:2021abf}
N.~Brambilla, H.~S.~Chung and A.~Vairo,
{\it Inclusive production of heavy quarkonia in pNRQCD},
JHEP \textbf{09} (2021) 032
[arXiv:2106.09417 [hep-ph]].

\bibitem{Bodwin:1994jh}
G.~T.~Bodwin, E.~Braaten and G.~P.~Lepage,
{\it Rigorous QCD analysis of inclusive annihilation and production of heavy quarkonium},
Phys. Rev. D \textbf{51} (1995) 1125-1171
[erratum: Phys. Rev. D \textbf{55} (1997) 5853]
[arXiv:hep-ph/9407339 [hep-ph]].

\bibitem{Nayak:2005rw}
G.~C.~Nayak, J.~W.~Qiu and G.~F.~Sterman,
{\it Fragmentation, factorization and infrared poles in heavy quarkonium production},
Phys. Lett. B \textbf{613} (2005) 45-51
[arXiv:hep-ph/0501235 [hep-ph]].

\bibitem{Nayak:2005rt}
G.~C.~Nayak, J.~W.~Qiu and G.~F.~Sterman,
{\it Fragmentation, NRQCD and NNLO factorization analysis in heavy quarkonium production},
Phys. Rev. D \textbf{72} (2005) 114012
[arXiv:hep-ph/0509021 [hep-ph]].

\bibitem{Nayak:2006fm}
G.~C.~Nayak, J.~W.~Qiu and G.~F.~Sterman,
{\it NRQCD factorization and velocity-dependence of NNLO poles in heavy quarkonium production},
Phys. Rev. D \textbf{74} (2006) 074007
[arXiv:hep-ph/0608066 [hep-ph]].

\bibitem{Bodwin:2019bpf}
G.~T.~Bodwin, H.~S.~Chung, J.~H.~Ee, U.~R.~Kim and J.~Lee,
{\it Covariant calculation of a two-loop test of nonrelativistic QCD factorization},
Phys. Rev. D \textbf{101} (2020) no.9, 096011
[arXiv:1910.05497 [hep-ph]].

\bibitem{Zhang:2020atv}
P.~Zhang, C.~Meng, Y.~Q.~Ma and K.~T.~Chao,
{\it Gluon fragmentation into $^{3} {P}_J^{\left[1,8\right]} $ quark pair and test of NRQCD factorization at two-loop level},
JHEP \textbf{08} (2021) 111
[arXiv:2011.04905 [hep-ph]].

\bibitem{Chung:2018lyq}
H.~S.~Chung,
{\it Review of quarkonium production: status and prospects},
PoS \textbf{Confinement2018} (2018) 007
[arXiv:1811.12098 [hep-ph]].

\bibitem{Lansberg:2019adr}
J.~P.~Lansberg,
{\it New observables in inclusive production of quarkonia},
Phys. Rept. \textbf{889} (2020) 1-106
[arXiv:1903.09185 [hep-ph]].

\bibitem{Brambilla:1999xf}
N.~Brambilla, A.~Pineda, J.~Soto and A.~Vairo,
{\it Potential NRQCD: an effective theory for heavy quarkonium},
Nucl. Phys. B \textbf{566} (2000) 275
[arXiv:hep-ph/9907240 [hep-ph]].

\bibitem{Brambilla:2000gk}
N.~Brambilla, A.~Pineda, J.~Soto and A.~Vairo,
{\it The QCD potential at $O(1/m)$},
Phys. Rev. D \textbf{63} (2001) 014023
[arXiv:hep-ph/0002250 [hep-ph]].

\bibitem{Pineda:2000sz}
A.~Pineda and A.~Vairo,
{\it The QCD potential at $O(1/m^2)$: complete spin dependent and spin independent result},
Phys. Rev. D \textbf{63} (2001) 054007
[erratum: Phys. Rev. D \textbf{64} (2001) 039902]
[arXiv:hep-ph/0009145 [hep-ph]].

\bibitem{Brambilla:2003mu}
N.~Brambilla, A.~Pineda, J.~Soto and A.~Vairo,
{\it The $(m \Lambda_{\rm QCD})^{1/2}$ scale in heavy quarkonium},
Phys. Lett. B \textbf{580} (2004) 60-71
[arXiv:hep-ph/0307159 [hep-ph]].

\bibitem{Bali:2000vr}
G.~S.~Bali \text{et al.} [TXL and T(X)L],
{\it Static potentials and glueball masses from QCD simulations with Wilson sea quarks},
Phys. Rev. D \textbf{62} (2000) 054503
[arXiv:hep-lat/0003012 [hep-lat]].

\bibitem{Juge:2002br}
K.~J.~Juge, J.~Kuti and C.~Morningstar,
{\it Fine structure of the QCD string spectrum},
Phys. Rev. Lett. \textbf{90} (2003) 161601
[arXiv:hep-lat/0207004 [hep-lat]].

\bibitem{Capitani:2018rox}
S.~Capitani, O.~Philipsen, C.~Reisinger, C.~Riehl and M.~Wagner,
{\it Precision computation of hybrid static potentials in SU(3) lattice gauge theory},
Phys. Rev. D \textbf{99} (2019) no.3, 034502
[arXiv:1811.11046 [hep-lat]].

\bibitem{Makeenko:1979pb}
Y.~M.~Makeenko and A.~A.~Migdal,
{\it Exact equation for the loop average in multicolor QCD},
Phys. Lett. B \textbf{88} (1979) 135
[erratum: Phys. Lett. B \textbf{89} (1980), 437].

\bibitem{Witten:1979pi}
E.~Witten,
{\it The $1/N$ expansion in atomic and particle physics},
NATO Sci. Ser. B \textbf{59} (1980), 403-419.

\bibitem{Brambilla:2017kgw}
N.~Brambilla, W.~Chen, Y.~Jia, V.~Shtabovenko and A.~Vairo,
{\it Relativistic corrections to exclusive $\chi_{cJ} + \gamma$ production from $e^+ e^-$ annihilation},
Phys. Rev. D \textbf{97} (2018) no.9, 096001
[erratum: Phys. Rev. D \textbf{101} (2020) no.3, 039903]
[arXiv:1712.06165 [hep-ph]].

\bibitem{Brambilla:2001xy}
N.~Brambilla, D.~Eiras, A.~Pineda, J.~Soto and A.~Vairo,
{\it New predictions for inclusive heavy quarkonium P wave decays},
Phys. Rev. Lett. \textbf{88} (2002) 012003
[arXiv:hep-ph/0109130 [hep-ph]].

\bibitem{Brambilla:2002nu}
N.~Brambilla, D.~Eiras, A.~Pineda, J.~Soto and A.~Vairo,
{\it Inclusive decays of heavy quarkonium to light particles},
Phys. Rev. D \textbf{67} (2003) 034018
[arXiv:hep-ph/0208019 [hep-ph]].

\bibitem{ParticleDataGroup:2020ssz}
P.~A.~Zyla \text{et al.} [Particle Data Group],
{\it Review of Particle Physics},
PTEP \textbf{2020} (2020) no.8, 083C01.

\bibitem{Belle:2018jqa}
S.~Jia \text{et al.} [Belle],
{\it Observation of $e^+e^- \to \gamma \chi_{c1}$ and search for $e^+e^- \to \gamma \chi_{c0}, \gamma \chi_{c2},$ and $\gamma\eta_c$ at $\sqrt{s}$ near 10.6 GeV at Belle},
Phys. Rev. D \textbf{98} (2018) no.9, 092015
[arXiv:1810.10291 [hep-ex]].

\bibitem{CMS:2012qwg}
S.~Chatrchyan \text{et al.} [CMS],
{\it Measurement of the relative prompt production rate of $\chi_{c2}$ and $\chi_{c1}$ in $pp$ collisions at $\sqrt{s}=7$ TeV},
Eur. Phys. J. C \textbf{72} (2012) 2251
[arXiv:1210.0875 [hep-ex]].

\bibitem{ATLAS:2014ala}
G.~Aad \text{et al.} [ATLAS],
{\it Measurement of $\chi_{c1}$ and $\chi_{c2}$ production with $\sqrt{s}$ = 7 TeV $pp$ collisions at ATLAS},
JHEP \textbf{07} (2014) 154
[arXiv:1404.7035 [hep-ex]].

\bibitem{Bodwin:2014gia}
G.~T.~Bodwin, H.~S.~Chung, U.~R.~Kim and J.~Lee,
{\it Fragmentation contributions to $J/\psi$ production at the Tevatron and the LHC},
Phys. Rev. Lett. \textbf{113} (2014) no.2, 022001
[arXiv:1403.3612 [hep-ph]].

\bibitem{Bodwin:2015iua}
G.~T.~Bodwin, K.~T.~Chao, H.~S.~Chung, U.~R.~Kim, J.~Lee and Y.~Q.~Ma,
{\it Fragmentation contributions to hadroproduction of prompt $J/\psi$, $\chi_{cJ}$, and $\psi(2S)$ states},
Phys. Rev. D \textbf{93} (2016) no.3, 034041
[arXiv:1509.07904 [hep-ph]].

\bibitem{CMS:2019jas}
A.~M.~Sirunyan \text{et al.} [CMS],
{\it Constraints on the $\chi_\mathrm{c1}$ versus $\chi_\mathrm{c2}$ polarizations in proton-proton collisions at $\sqrt{s} =$ 8 TeV},
Phys. Rev. Lett. \textbf{124} (2020) no.16, 162002
[arXiv:1912.07706 [hep-ex]].

\bibitem{LHCb:2014nug}
R.~Aaij \text{et al.} [LHCb],
{\it Measurement of the $\chi_b(3P)$ mass and of the relative rate of $\chi_{b1}(1P)$ and $\chi_{b2}(1P)$ production},
JHEP \textbf{10} (2014) 088
[arXiv:1409.1408 [hep-ex]].

\bibitem{CMS:2014bsd}
V.~Khachatryan \text{et al.} [CMS],
{\it Measurement of the production cross section ratio $\sigma(\chi_{b2}(\mathrm{1P}))/\sigma(\chi_{b1}(\mathrm{1P}))$ in pp collisions at $\sqrt s $ = 8 TeV},
Phys. Lett. B \textbf{743} (2015) 383-402
[arXiv:1409.5761 [hep-ex]].

\bibitem{LHCb:2014ngh}
R.~Aaij \text{et al.} [LHCb],
{\it Study of $\chi _{{\mathrm {b}}}$ meson production in $pp$ collisions at $\sqrt{s}=7$ and $8{\mathrm {\,TeV}} $
  and observation of the decay $\chi _{{\mathrm {b}}}\mathrm {(3P)} \rightarrow \Upsilon \mathrm {(3S)} {\gamma } $},
Eur. Phys. J. C \textbf{74} (2014) no.10 3092
[arXiv:1407.7734 [hep-ex]].

\bibitem{Chung:2020zqc}
H.~S.~Chung,
{\it $\overline {MS}$ renormalization of $S$-wave quarkonium wavefunctions at the origin},
JHEP \textbf{12} (2020) 065
[arXiv:2007.01737 [hep-ph]].

\bibitem{Chung:2021efj}
H.~S.~Chung,
{\it $P$-wave quarkonium wavefunctions at the origin in the $\overline{MS}$ scheme},
[arXiv:2106.15514 [hep-ph]].

\bibitem{Brambilla:2019fmu}
N.~Brambilla, H.~S.~Chung, W.~K.~Lai, V.~Shtabovenko and A.~Vairo,
{\it Order $v^4$ corrections to Higgs boson decay into $J/\psi + \gamma$},
Phys. Rev. D \textbf{100} (2019) no.5, 054038
[arXiv:1907.06473 [hep-ph]].

\end{thebibliography}
\end{document}